\documentclass[12pt]{iopart}
\usepackage{iopams}
\usepackage{graphicx}

\eqnobysec
\begin{document}
\title[
Confinement interaction in nonlinear Wick-Cutkosky model]
{Confinement interaction in nonlinear generalizations of
the Wick-Cutkosky model}
\author{J W Darewych\dag ~and~ A Duviryak\ddag}
\address{\dag\
Department of Physics and Astronomy, York
University, Toronto, Ontario, M3J~1P3, Canada}
\address{\ddag\
Department for Computer Simulations of Many-Particle Systems,
Institute for
Condensed Matter Physics of NAS of Ukraine, Lviv, UA-79011, Ukraine
}
\ead{\dag\ darewych@yorku.ca, \ddag\ duviryak@ph.icmp.lviv.ua}

\newcommand{\ha}{\mbox{\small$\frac{1}{2}$}}
\renewcommand{\th}{\mbox{\small$\frac{1}{3}$}}
\newcommand{\qu}{\mbox{\small$\frac{1}{4}$}}
\newcommand{\lab}[1]{\label{#1}}
\newcommand{\re}[1]{(\ref{#1})}
\newcommand{\nn}{\nonumber}
\newcommand{\B}[1]{\bi{#1}}
\newcommand{\Blambda}{\mbox{\boldmath$\lambda$}}
\newcommand{\st}[2]{\lefteqn{\stackrel{\smash{#1}}{#2}}\phantom{#2}}
\newcommand{\D}[2]{{\rm d}^{#1}{#2}\,}
\newcommand{\inta}{\hspace*{-.1cm}\int \hspace*{-.15cm}}
\newcommand{\intabc}{\hspace*{-.1cm}\int\hspace*{-.2cm}\int\hspace*{-.2cm}\int
                    \hspace{-.15cm}}
\def\ds{\displaystyle}
\def\Ub{{\bar U}}

\begin{abstract}
We consider nonlinear-mediating-field generalizations of
the Wick-Cutkosky model. Using an iterative approach and eliminating
the mediating field by means of the covariant Green function we
arrive at a Lagrangian density containing many-point time-nonlocal
interaction terms. In low-order approximations of
$\varphi^3{+}\varphi^4$ theory we obtain the usual
two-current interaction as well as a three-current interaction of a
confining type. The same result is obtained without approximation
for a version of the dipole model. The transition to the Hamiltonian
formalism and subsequent canonical quantization is performed with
time non-locality taken into account approximately.

    A relativistic three-particle wave equation is derived
variationally by using  a three-particle Fock space trial state.
The non-relativistic limit of this equation is
obtained and its properties are analyzed and discussed.
\end{abstract}

\pacs{11.10Ef, 11.10Lm}

\submitto{\JPA}


\maketitle


\section{Introduction}

    Confinement is evidently related to
the nonlinearity of chromodynamics. Since confining solutions
of classical non-Abelian field equations are not known at present
\cite{Act79}, it is believed that confinement is an
essentially quantum effect. This is supported
by numerical computations of QCD on the
lattice \cite{Gre03,Swa04}. However, the analytical
study of confinement, particularly in gauge field theory like QCD,
remains a challenging task \cite{Swa04}. Thus the  study
of simpler field-theoretical models that simulate the
characteristic features of confinement remains relevant.

    In this regard, it is worth mentioning earlier models, such as
the dipole model \cite{Bla74} and the related higher derivative
model \cite{Kis75} with its subsequent non-Abelian generalization
\cite{A-A84}. They indicate a $1/k^4$ infrared behavior of the
``gluon" propagator, and thus a linear interaction potential, even
at the classical level. In spite of some quantization
inconsistencies, these phenomenological models treat the confinement
interaction as an elementary process, i.e., a two-particle
interaction arising from the lowest-order approximation of
perturbative dynamics of the models.

   More realistic models are the Dual Abelian Higgs model \cite{Swa04}
and non-Abelian versions \cite{K-S00,Swa04} in which the spontaneous
symmetry breaking mechanism is used to generate a vacuum condensate
with confining properties. In this approach the confinement interaction
is a kind of collective effect similar to that in condensed matter physics.

    The two classes of models mentioned above represent quite different
points of view on the confinement mechanism. The purpose of the
present study is to shed light on the question: is an intermediate
mechanism possible, in which confinement cannot be reduced to an
elementary processes but is governed by cluster interactions
involving finite numbers of particles?

    To investigate such a possibility, we utilize
the variational method, in a reformulated Hamiltonian formalism of
Quantum Field Theory (QFT), which has been demonstrated to be a
promising and powerful approach to the relativistic bound state
problem \cite{Dar98}--\cite{ERD06}. In particular, this approach has
been used to derive (and solve approximately) relativistic equations
for two and three fermion systems, such as Positronium (Ps) and
Muonium (Mu) \cite{TD1},  and also Ps$^-$ and Mu$^-$ \cite{BD}, and
it was shown that the derived bound state energies agree with
conventional perturbation theory and with experimental results
(where available).

    The use of many-particle Fock-space components in the variational
trial states leads to wave equations with systematically improvable
bound state energy levels, as has been shown, for example, on the
simple scalar Yukawa model \cite{ERD05,ERD06}.

    In this paper, we analyze the interactions that arise from the
non-linear terms in the mediating-field sector of the QFT
Lagrangian. In particular we consider the
$(\varphi^3{+}\varphi^4)$-generalization of the Wick-Cutkosky (i.e.
massless scalar Yukawa) model \cite{WC} as well as a version of the
dipole model \cite{Bla74,D-D04}.

We note that the models being considered are not of a
non-Abelian gauge-field type. The only two features which are common
to the models of this paper and QCD are the massless and non-linear
nature of the mediating field. Both features are important in the
generating confinement but the mechanism of this effect here is
different from that in gauge models \cite{Swa04,K-S00}.


\section{Partially reduced Wick-Cutkosky model}

    The Wick-Cutkosky model \cite{WC} is based on the classical action integral:
%
\begin{equation}\lab{2.1}
I=\int \D4x {\cal L}(x),
\end{equation}
with the Lagrangian density ($\hbar = c =1$)
%
\begin{equation}\lab{2.2}
{\cal L} =\partial_\mu\phi^*\partial^\mu\phi - m^2\phi^*\phi -g\phi^*\phi\, \chi
+ \ha\partial_\mu\chi\partial^\mu\chi,
\end{equation}
where $\phi(x)$ is a complex scalar ``matter" field with rest mass
$m$, and $\chi(x)$ is a real massless scalar field interacting with
$\phi$ via the Yukawa term $g\phi^*\phi\, \chi$ (here $g$ is
an interaction constant).

    The stationary property of the action \re{2.1}-\re{2.2}, i.e.
$\delta I= 0$, leads to the coupled set of the Euler-Lagrange equations,
%
\begin{eqnarray}
& (\square + m^2)\phi =-g \phi\chi, \lab{2.3}\\
& (\square + m^2)\phi^* = - g\phi^*\chi, \lab{2.4}\\
& \square\chi = \rho, \lab{2.5}
\end{eqnarray}
which determine the field dynamics; here $\rho\equiv-g\phi^*\phi$.

    Equation \re{2.5} can be solved exactly:
%
\begin{equation}\lab{2.6}
\chi = D*\rho+\chi_0,
\end{equation}
where ``~$*$~'' denotes the convolution
$\left[D*\rho\right](x)\equiv\int\D4x'D(x-x')\rho(x')$ and
$D(x)=\frac1{4\pi}\delta(x^2)$ is the symmetric Green function of
the d'Alembert equation. If the free $\chi$ field plays no role in
the investigation the arbitrary solution $\chi_0$ of the homogeneous
d'Alembert equation can be omitted. Then the use of the formal
solution \re{2.6} (with $\chi_0=0$) in the Lagrangian \re{2.2} leads
to a self-contained variational principle for the interacting fields
$\phi(x)$ and $\phi^*(x)$. The modified Lagrangian $\bar{\cal L}$
which we shall refer to as the partially-reduced Lagrangian, is an
important basis for the quantization of the model; cf. refs.
\cite{Dar00,D-D04}.

We demonstrate here how to derive the partially-reduced Lagrangian
for the Wick-Cutkosky model without the use of the condition $\chi_0=0$. For
this purpose we consider the equality \re{2.6} as a change of
variable $\chi\to\chi_0$ where the new field $\chi_0$ is not {\em
a'priori} subjected to any field equation. The substitution of
\re{2.6} directly in the Lagrangian \re{2.2} gives
%
\begin{eqnarray}\lab{2.7}
{\cal L} &=&\partial_\mu\phi^*\partial^\mu\phi - m^2\phi^*\phi +
\rho(D*\rho+\chi_0) +
\ha\left[\partial_\mu(D*\rho+\chi_0)\partial^\mu(D*\rho+\chi_0)\right]\nn\\
&\simeq&
\partial_\mu\phi^*\partial^\mu\phi - m^2\phi^*\phi +
\rho(D*\rho+\chi_0) -
\ha(D*\rho+\chi_0)\square(D*\rho+\chi_0)\nn\\
&\simeq& \underbrace{\partial_\mu\phi^*\partial^\mu\phi -
m^2\phi^*\phi + \ha\rho D*\rho}_{\displaystyle{\bar{\cal L}}} +
\underbrace{\ha\partial_\mu\chi_0\partial^\mu\chi_0}_{\displaystyle{\Delta{\cal
L}_{\mathrm{free}}}},
\end{eqnarray}
where $\simeq$ denotes equality modulo surface terms. In this form
the system is effectively split into two independent subsystems: the
interacting $\phi$ matter field and the free $\chi_0$ field. From
this point on the physically trivial $\chi_0$-dependent $\Delta{\cal
L}_{\mathrm{free}}$ term can be ignored (as indicated above).
\footnote{ It is noteworthy that, within the variational problem
based on \re{2.7}, the primary meaning of $\chi_0$ in \re{2.6} as
general solution of the homogeneous d'Alembert equation is
restored.}

    The partially-reduced Lagrangian $\bar{\cal L}$
is non-local in space-time coordinates. The treatment of non-local
theories of this type is a conceptually intricate, but practically
realisable procedure. In particular, partially-reduced versions of
Yukawa-like models are worked out in \cite{D-D04}. In the next
section we consider a non-linear generalization of Wick-Cutkosky
model within the partially-reduced formulation.


\section{Nonlocal Lagrangian from a nonlinear Wick-Cutkosky model}

    We proceed from the Lagrangian density
%
\begin{equation}\lab{3.1}
{\cal L} =\partial_\mu\phi^*\partial^\mu\phi - m^2\phi^*\phi -g\phi^*\phi\, \chi
-\qu\lambda(\phi^*\phi)^2
+ \ha\partial_\mu\chi\partial^\mu\chi - {\cal V}(\chi),
\end{equation}
where $\lambda>0$ is a self-interaction coupling constant and ${\cal
V}(\chi)$ is an arbitrary potential (all other quantities are the
same as in \re{2.2}).

    The new terms, $\lambda(\phi^*\phi)^2$ and ${\cal V}(\chi)$,
modify the Euler-Lagrange equations  \re{2.3}-\re{2.5}. In
particular, the equation \re{2.5} becomes the non-linear
inhomogeneous d'Alembert equation
%
\begin{equation}\lab{3.2}
\square\chi = \rho - {\cal V}\,'(\chi),
\end{equation}
where ${\cal V}\,'(\chi)\equiv{\rm d{\cal V}(\chi)}/{\rm d}\chi$.
It can be formally solved by iteration
(cf. ref. \cite{ShpDar02}). In the 1st-order approximation
we have:
\begin{equation}\lab{3.3}
\chi = D*[\rho-{\cal V}\,'(D*\rho)]+\chi_0,
\end{equation}
where $\chi_0$ includes an
arbitrary solution of the homogeneous equation.

    Similarly to the case of the linear Wick-Cutkosky model, we use the
replacement \re{3.3} (where $\chi_0$ is a new field variable)
in the Lagrangian \re{3.1}. In 1st order this gives,
%
\begin{eqnarray}\lab{3.4}
{\cal L} &\simeq&
\partial_\mu\phi^*\partial^\mu\phi - m^2\phi^*\phi +
\ha\rho D*\rho -\qu\lambda(\phi^*\phi)^2\nn\\
&&{}+\ha\partial_\mu\chi_0\partial^\mu\chi_0
-{\cal V}(D*\rho+\chi_0)+\chi_0{\cal V}'(D*\rho).
\end{eqnarray}
Unlike the Lagrangian \re{2.7}, this functional is not
completely split in the $\phi$ and $\chi_0$ variables. The Euler-Lagrange
equation for $\chi_0$,
%
\begin{equation}\lab{3.5}
\square\chi_0 = -{\cal V}'(D*\rho+\chi_0) + {\cal V}'(D*\rho),
\end{equation}
is a free-field one only in zero-order approximation. Nevertheless, it possesses
the solution $\chi_0=0$ which, upon substitution into \re{3.4}, gives the
reduced Lagrangian:
%
\begin{eqnarray}\lab{3.6}
\bar{\cal L} &\simeq
\partial_\mu\phi^*\partial^\mu\phi - m^2\phi^*\phi +
\ha\rho D*\rho -\qu\lambda(\phi^*\phi)^2-{\cal V}(D*\rho)\nn\\
&
\equiv{\cal L}_{\rm free} + {\cal L}_{\rm int}^{(2)} + {\cal
L}_{\rm int}^{(>2)}
\end{eqnarray}
It is non-local, and the action \re{2.1},
\re{3.6} includes 1-, 2- and {$>$$2$\,}-fold integrations over
Minkowsky space.

The difference $\Delta{\cal L}={\cal L}-\bar{\cal L}$, i.e.,
the $\chi_0$-dependent part of the total Lagrangian \re{3.4},
is at least quadratic in  the $\chi_0$ variable:
%
\begin{eqnarray}\lab{3.7}
\Delta{\cal L} &= \Delta{\cal L}_{\rm free} + \Delta{\cal L}_{\rm int} \quad\mbox{where}\quad
\Delta{\cal L}_{\rm free}=\ha\partial_\mu\chi_0\partial^\mu\chi_0,\nn\\
\Delta{\cal L}_{\rm int}
&
={\cal V}(D*\rho)+\chi_0{\cal V}'(D*\rho)-{\cal V}(D*\rho+\chi_0)\nn\\
\fl&={}-\frac1{2!}\chi_0^2{\cal V}''(D*\rho)-\frac1{3!}\chi_0^3{\cal
V}'''(D*\rho)-\dots
\end{eqnarray}
{This structure shows that the term $\Delta{\cal L}$ is not
important in the present work, as will be explained}
in more detail in Section 7.

    The non-local Lagrangian \re{3.6} is the 1st-order approximate
result of the reduction procedure applied to nonlinear
generalizations of the Wick-Cutkosky model. In the Appendix we
construct another local model, a kind of dipole model (with a pair
of mediating fields), that can be reduced to the Lagrangian \re{3.6}
exactly.


\section{Quantization}

    In order to proceed farther we need to specify the interaction potential
${\cal V}(\chi)$. We choose
%
\begin{equation}\lab{4.1}
{\cal V}(\chi)=\th\kappa\chi^3 + \qu\varkappa\chi^4,
\end{equation}
where $\kappa$ and $\varkappa > 0$ are coupling constants. In this
case the nonlinear Wick-Cutkosky model \re{3.1} possesses a stable
perturbative vacuum and is renormalizable.

We proceed from the partially reduced Lagrangian \re{3.6}, construct
the Hamiltonian of the model and perform the canonical
quantization. Due to the non-locality of the Lagrangian \re{3.6},
the Hamiltonization is a rather complicated procedure. It can be
performed perturbatively, following Refs.
\cite{D-D04}--\cite{ERD06}. In leading-order approximation the
Hamiltonization proceeds as follows. We work out the Hamiltonian
density,
%
\begin{equation}\lab{4.2}
{\cal H}={\cal H}_{\rm free}+{\cal H}_{\rm int}^{(2)}
+{\cal H}_{\rm int}^{(3)}+{\cal H}_{\rm int}^{(4)},
\end{equation}
where
%
\begin{eqnarray}
\fl
{\cal H}_{\rm int}^{(2)}(x) &= -\ha\int \D4{x'} \rho(x) D(x-x') \rho(x') +
\qu\lambda\left(\phi^*(x)\phi(x)\right)^2\nn\\
\fl&
\equiv-\ha\int \D4{x'} \rho(x)\left[D(x-x')-\frac\lambda{2g^2}\delta(x-x')\right]\rho(x'),
\lab{4.3}\\
\fl
{\cal H}_{\rm int}^{(3)} (x) &=
\th\kappa\intabc\,\D4x'\D4x''\D4zD(z-x)D(z-x')D(z-x'')\rho(x)\rho(x')\rho(x''),
\lab{4.4}\\
\fl
{\cal H}_{\rm int}^{(4)} (x) &=
\qu\varkappa\inta\intabc\,\D4x'\D4x''\D4x'''\D4zD(z-x)D(z-x')D(z-x'')D(z-x''')\times\nn\\
\fl&\hspace{21em}
{}\times\rho(x)\rho(x')\rho(x'')\rho(x''').
\lab{4.5}
\end{eqnarray}
The total interaction Hamiltonian density \re{4.2} is then expressed
in terms of the Fourier amplitudes $A_{\bi k}$, $B_{\bi k}$ and
$A^\dag_{\bi k}$, $B^\dag_{\bi k}$, of the field $\phi(x)$ (see eq.
(2.14) in \cite{ERD06}; actually, the procedure is somewhat more
intricate \cite{D-D04} but the result is the same). Upon
quantization these amplitudes satisfy the standard commutation
relations and become the creation and annihilation operators. Then
the canonical Hamiltonian operator is given by
%
\begin{equation}\lab{4.6}
H=\int\D3x:{\cal H}(t{=}0,\B x):  \, ,
\end{equation}
where ``:\quad:" denotes the normal ordering of operators.
Other canonical generators, such as linear and angular momentum,
can be easily obtained.

    The term $H_{\rm free}$ is the standard Hamiltonian of the free complex scalar field.
The explicit form of the pair interaction term $H_{\rm int}^{(2)}$
is known (see \cite{Dar00,D-D04}) and
so we shall concentrate on the $H_{\rm int}^{(3)}$ term.
It has the following somewhat cumbersome form:
%
\begin{eqnarray}
\fl
H_{\rm int}^{(3)}= -\frac{\kappa
g^3}{24(2\pi)^6}\int\frac{\D3k_1\dots\D3k_6}{\sqrt{k_{10}\dots
k_{60}}} \sum\limits_{{\eta_1{=}\pm\atop{\smash{\dots\dots\atop\eta_6{=}\pm}}}}
\tilde D(\eta_1k_1+\eta_2k_2)\tilde D(\eta_3k_3+\eta_4k_4)\tilde D(\eta_5k_5+\eta_6k_6)\times\nn\\
\fl\hspace{23ex}
{}\times \delta(\eta_1\B k_1+\dots+\eta_6\B
k_6):\stackrel{\eta_1}B_{{\B k}_1}\stackrel{\eta_2}A_{{\B k}_2}
\stackrel{\eta_3}B_{{\B k}_3}\stackrel{\eta_4}A_{{\B
k}_4}\stackrel{\eta_5}B_{{\B k}_5} \stackrel{\eta_6}A_{{\B k}_6}:\, ,
\lab{4.7}
\end{eqnarray}
where ${\stackrel{+}B}=B$, ${\stackrel{-}B}=A^\dag$,
${\stackrel{+}A}=A$, ${\stackrel{-}A}=B^\dag$ and the Fourier
transform, $\tilde D(k)=-{\cal P}/k^2$, of the symmetric Green
function of the d'Alembert  equation depends on the on-shell 4-momentum
$k=\{k_0,\B k\}$, where $k_0=\sqrt{m^2+\B k^2}$. The expression
\re{4.7}
includes $2^6=64$ terms. The term $H_{\rm int}^{(4)}$
is of similar but more cumbersome form. We do not exhibit it
explicitly, since, as will be seen below, it makes no contribution
to the three-body equation derived in this work.


\section{Variational three-particle wave equations}

    In the variational approach to QFT the trial state of the system is built of few particle channel
components \cite{ERD05,ERD06} such as the two-particle state vector
$|2\rangle = \frac1{\sqrt{2}}\inta\D3{p_1}\D3{p_2} \, F_2(\B p_1,\B p_3)
\, A_{{\B p}_1}^\dag A_{{\B p}_2}^\dag |0\rangle$,
the particle-antiparticle one $|1+\bar1\rangle = \inta\D3{p_1}\D3{p_2} \, G(\B p_1,\B p_3)
\, A_{{\B p}_1}^\dag B_{{\B p}_2}^\dag |0\rangle$, and so
on. The three-particle component has the form
%
\begin{equation}\lab{5.1}
|3\rangle=\frac1{\sqrt{3!}}\int\D3{p_1}\D3{p_2}\D3{p_3} \, F(\B p_1,\B p_2,\B p_3)
\, A_{{\B p}_1}^\dag A_{{\B p}_2}^\dag A_{{\B p}_3}^\dag|0\rangle,
\end{equation}
where the channel wave function $F$,
which is to be determined variationally,
is completely symmetric under
the permutation of the variables $\B p_1,\B p_2,\B p_3$.
In the variational method the channel components, $|\psi_N\rangle$,
are used to determine the matrix elements of the Hamiltonian, namely
$\langle \psi_N| H | \psi_{N'}\rangle$,
where $N, N'$ stand for $1, \,{\bar 1}, \, 2, \, 1{+}{\bar 1},
\, {\bar 2}, \, 3,\, 2{+}{\bar 1},\, 2{+}{\bar 2}, \dots$

    We are interested here in the matrix element of the interaction
$H_{\rm int}=H_{\rm int}^{(2)}+H_{\rm int}^{(3)}+H_{\rm int}^{(4)}$
of the Hamiltonian. We note that
${\langle1{+}\bar1|H_{\rm int}^{(3)}|1{+}\bar1\rangle}=0$,
${\langle2|H_{\rm int}^{(3)}|2\rangle}=0$. In other words, purely
two-particle trial states, and so the resulting
variational wave equations,
do not sample the term $H_{\rm int}^{(3)}$.
Thus we first consider the three-particle case and calculate the
matrix element
%
\begin{eqnarray}
\fl
\langle3|H_{\rm int}|3\rangle=\inta\D3{p'_1}...\D3{p'_3} \, \D3p_1...\D3p_3 \,
{F^*}(\B p'_1...\B p'_3) \, F(\B p_1...\B p_3)\,
{\cal K}_{33}(\B p'_1...\B p'_3,\B p_1...\B p_3) ,
\lab{5.2}
\end{eqnarray}
where the kernel ${\cal K}_{33}={\cal K}_{33}^{(2)}+{\cal K}_{33}^{(3)}$
consists of the following components:
%
\begin{eqnarray}
\fl
{\cal K}_{33}^{(2)}(\B p'_1...\B p'_3,\B p_1...\B p_3)= &-\frac{3}{4(2\pi)^3}
\, \delta(\B p'_1+\B p'_2+\B p'_3-\B p_1-\B p_2-\B p_3)\times\nn\\
&\times
\frac{\delta(\B p'_3-\B p_3)}
{\sqrt{p'_{10}p'_{20}p_{10}p_{20}}}
\left[g^2\tilde D(p'_2-p_2)-\lambda/2\right],
\lab{5.3}\\
\fl
{\cal K}_{33}^{(3)}(\B p'_1...\B p'_3,\B p_1...\B p_3)=&
-\frac{\kappa g^3}{4(2\pi)^6} \,
\delta(\B p'_1+\B p'_2+\B p'_3-\B p_1-\B p_2-\B p_3)\times\nn\\
&\times
\frac{\tilde D(p'_1-p_1)\tilde D(p'_2-p_2)\tilde D(p'_3-p_3)}
{\sqrt{p'_{10}...p'_{30}p_{10}...p_{30}}},
\lab{5.4}
\end{eqnarray}
and  $p_{i0} = \sqrt{m^2 +{\B p}_i^2}$ and similarly for $p'_{j0}$ ($i,j=1,2,3$).
The term $H_{\rm int}^{(4)}$ does not contribute in ${\cal K}_{33}$,
i.e., ${\cal K}_{33}^{(4)}=0$.

    The kernel ${\cal K}_{33}$ determines the interaction in the relativistic
three-particle  wave equation that follows from the variational
principle $\delta \, \langle 3 | H - E| 3 \rangle = 0$, namely
%
\begin{eqnarray}\lab{5.5}
\fl
\{p_{10}+p_{20}+p_{30}&-E\} F(\B p_1,\B p_2, \B p_3) \nn\\
\fl&{}+ \inta \D3{p'_1}\D3{p'_2}\D3{p'_3} \,
{\cal K}_{33}(\B p_1,\B p_2,\B p_3,\B p'_1, \B p'_2,\B p'_3)
\, F(\B p'_1, \B p'_2, \B p'_3)=0
\end{eqnarray}
where the kernel is understood
to be the completely symmetrized expression (with respect to
the variables $\B p'_1,\B p'_2,\B p'_3$ and $\B p_1,\B p_2,\B p_3$) of
\re{5.3} and \re{5.4}.

    The term ${\cal K}_{33}^{(2)}$ of the kernel corresponds to the attractive interaction
via massless boson exchange and repulsive contact interaction between each
pair of particles while ${\cal K}_{33}^{(3)}$ describes a cluster three-particle interaction.

From the mathematical viewpoint the three-body wave-equation \re{5.5} is
an integral equation
with a singular kernel. Even in simpler (say, two-particle) cases such equations are usually
solved approximately (variationally, numerically, perturbatively), and it is
not easy to get a general qualitative
characteristic of the solutions, or to estimate the role of different terms of the kernel.

    In order to have some understanding of the properties
of the cluster interaction we consider the non-relativistic limit of the
equation \re{5.5}, in which
case the kernels simplify considerably, and then perform
the Fourier transformation into
coordinate space. In this representation the equation
is simply a Schr\"odinger equation for the three-particle eigenfunction
$\Psi(\B x_1,\B x_2,\B x_3)$
(see \cite{ERD05}) and eigenenergy $\epsilon=E-3m$:
%
\begin{eqnarray}
\left\{\frac1{2m}(\B p_1^2 + \B p_2^2 +\B p_3^2) + V(\B x_1,\B x_2,\B x_3) -
\epsilon\right\}\Psi(\B x_1,\B x_2,\B x_3)=0,
\lab{5.6}
\end{eqnarray}
where $\B p_i=-\rmi\bnabla_i$ ($i=1,2,3$), and the potential $V(\B x_1,\B x_2,\B x_3)$,
like the relativistic kernel ${\cal K}_{33}$,
consists of two parts, $V=V_{33}^{(2)}+V_{33}^{(3)}$:
%
\begin{eqnarray}
V_{33}^{(2)}(\B x_1,\B x_2,\B x_3)=&-\frac{g^2}{16\pi m^2}\left\{\frac1{|\B x_1-\B x_2|}+\frac1{|\B x_2-\B x_3|}+
\frac1{|\B x_3-\B x_1|}\right\}\nn\\
&{}+\frac{\lambda}{8m^2}\left\{\delta(\B x_1-\B x_2)+\delta(\B x_2-\B x_3)+
\delta(\B x_3-\B x_1)\right\},
\lab{5.7}\\
V_{33}^{(3)}(\B x_1,\B x_2,\B x_3)
=&\frac{2\kappa g^3}{(8\pi m)^3}U(\B x_1,\B x_2,\B x_3).
\quad
\lab{5.8}
\end{eqnarray}
where
%
\begin{equation}\lab{5.9}
U(\B x_1,\B x_2,\B x_3)\equiv-\!\int\!\frac{\D3z}
{|\B z-\B x_1||\B z-\B x_2||\B z-\B x_3|} .
\end{equation}
The integral in r.h.s. of \re{5.9}
is a divergent quantity and thus equation
\re{5.6} may seem to be meaningless. However,
the gradients $\partial
U(\B x_1, \B x_2, \B x_3)/\partial\B x_i$ ($i=1,2,3$) which determine
the forces in the classical background of this problem, are
well defined and finite. Thus the ``function"  \re{5.9}
can be presented in the form
%
\begin{equation}\lab{5.10}
U(\B x_1,\B x_2,\B x_3)=\tilde U(\B x_1,\B x_2,\B x_3)+U_0
\end{equation}
where $\tilde U(\B x_1,\B x_2,\B x_3)$ in a regular (finite) function and
$U_0$ is an infinite negative constant (independent of
the variables $\B x_1$, $\B x_2$, $\B x_3$).
This constant
can be absorbed by the eigenenergy $\epsilon$ so that
the wave equation \re{5.6} gets reformulated as follows:
%
\begin{eqnarray}
V_{33}^{(3)}(\B x_1,\B x_2,\B x_3)\to\tilde V_{33}^{(3)}(\B x_1,\B x_2,\B x_3)&=
\frac{2\kappa g^3}{(8\pi m)^3}\{U(\B x_1,\B x_2,\B x_3)-U_0\} \nn\\
&\equiv
\frac{2\kappa g^3}{(8\pi m)^3}\tilde U(\B x_1,\B x_2,\B x_3),
\lab{5.11}\\
\epsilon\to\tilde{\epsilon}=E-3m-\frac{2\kappa g^3}{(8\pi m)^3}U_0
\lab{5.12}
\end{eqnarray}
where the eigenenergy $\tilde\epsilon$ is finite (as
is the potential $\tilde V_{33}^{(3)}$).

In order to perform this reformulation explicitly, we need to resort to regularization
of the integral \re{5.9} which we consider in the next section.

    The problem of divergences is expected in the relativistic case too. But
the analysis of the integral equation \re{5.5}
is a more subtle problem which shall not be undertaken in this work.


\section{Properties and evaluation of the 3-point potential}

    Various regularization procedures are possible. In essence, one introduces
some cut-off parameter which finally is put to 0 (or $\infty$).
We enumerate some possibilities:
\medskip

\noindent
1) We could consider the case where the mediating $\chi$ field is massive, whereupon there would be a
mass term $- \ha \, \mu^2 \, \chi^2$ in the Lagrangian \re{3.1}. In that case the gravity-like
$~\ds \frac{1}{r}~$ factors would be replaced by the Yukawa forms $\ds \frac{e^{-\mu r}}{r}$. Thus
we could regard $U$ of eq. \re{5.9} as the massless-mediating-field limit of the
massive-mediating-field case,
%
\begin{equation} \lab{6.1}
U_\mu(\B x_1,\B x_2,\B x_3) = - \int\! \D3z \frac{e^{-\mu |\B z-\B x_1|}}{|\B z-\B x_1|} \,
\frac{e^{-\mu |\B z-\B x_2|}}{|\B z-\B x_2|} \, \frac{e^{-\mu |\B z-\B x_3|}}{|\B z-\B x_3|},
\end{equation}
which is well defined and finite for any $\mu > 0$.

    We note that by changing the variable of integration from $\B z$ to $\B v = \B z - \B x_1$
in eq. \re{6.1}, we can write $U_\mu$ as
%
\begin{equation} \lab{6.2}
U_\mu(\B x_1,\B x_2,\B x_3) = - \int\! \D3v \frac{e^{-\mu v}}{v} \,
\frac{e^{-\mu |\B v + \B x_{12}|}}{|\B v + \B x_{12}|} \, \frac{e^{-\mu |\B v + \B x_{13}|}}{|\B v+ \B x_{13}|}
 = \Ub_\mu(\B x_{12}, \B x_{13}),
\end{equation}
where $\B x_{ij} = \B x_i - \B x_j$ and $v = |\B v|$.
\medskip

\noindent
2) Another way would be to regard $U$ of eq. \re{5.9} as a limiting case, as $R \to \infty$, of
%
\begin{equation}  \lab{6.3}
\Ub_R(\B x_{12}, \B x_{13}) = - \int_0^R \D{}v \, v \int \D{}{\hat{\B v}} \, \frac{1}
{|\B v + \B x_{12}||\B v +\B x_{13}|},
\end{equation}
where $\hat{\B v} = {\B v}/v$ and  $R$ is an arbitrarily large, but finite, ``radius of space".
\medskip

\noindent
3) We could, also, regard $U$ of eq. \re{5.9} as a limiting case, as $\Lambda \to 0_+$, of
%
\begin{equation}  \lab{6.4}
\Ub_\Lambda (\B x_{12}, \B x_{13}) = - \int_0^\infty \D{}v \, v \, e^{- \Lambda v} \int \D{}{\hat{\B v}} \, \frac{1}
{|\B v + \B x_{12}||\B v+\B x_{13}|}.
\end{equation}
Evidently, any other suitable and convenient  cut-off function can be used
in place of $e^{-\Lambda v}$.

Of course, physical results would be meaningful to the extent that they
were independent of the choice of the regularization procedure.

Below we establish some general properties of
the regularized $\tilde U$ function. We discuss in
detail a convenient method of its evaluation
and show that it possesses a logarithmic confining property when $\mu \to 0$.

    Let us consider the regularization $U_\mu(\B x_1,\B x_2,\B x_3)$ \re{6.1}.
It obviously obeys the following symmetry properties:
\begin{enumerate}
\item
translational invariance: $U_\mu(\B x_1+\Blambda, \B x_2+\Blambda,
\B x_3+\Blambda) = U_\mu(\B x_1, \B x_2, \B x_3)$, where
$\Blambda\in\Bbb R^3$;
\item
rotational invariance: $U_\mu({\rm R}\B x_1, {\rm R}\B x_2, {\rm R}\B x_3) =
U_\mu(\B x_1, \B x_2, \B x_3)$, where ${\rm R}\in{\rm SO(3)}$;
\item
permutational invariance: $U_\mu(\B x_2, \B x_1, \B x_3 ) =U_\mu(\B x_1, \B x_3, \B
x_2 ) = U_\mu(\B x_1, \B x_2, \B x_3)$;
\item
scaling transformation: $U_\mu(\lambda\B x_1, \lambda\B x_2, \lambda\B x_3) =
U_{\lambda\mu}(\B x_1, \B x_2, \B x_3)$, where $\lambda\in\Bbb R_+$.
\end{enumerate}
These properties have implications for the structure of the regularized potential.

The properties (i)--(iii) hold for arbitrary
values of the cut-off parameter $\mu$, including the
formal limiting case $\mu\to0$.
Moreover, these are fundamental symmetries inherent to
any interaction potential of a
closed (nonrelativistic) system of three identical particles.
Thus the regularized potential
must possess the properties (i)--(iii) of necessity.

The scaling property (iv) has specific implication for the regularization \re{6.1}. In the
formal limit $\mu\to0$ the ``function'' $U\equiv U_{\mu=0}$ is scale invariant:
\medskip

\noindent
(iv)'
scale invariance: $U(\lambda\B x_1, \lambda\B x_2, \lambda\B x_3) =
U(\B x_1, \B x_2, \B x_3)$, where $\lambda\in\Bbb R_+$.
\medskip

\noindent
However, as is shown below, the scaling property of the regularized potential
$\tilde U(\B x_1, \B x_2, \B x_3)$ is different.

We  note that an important property of the potential $U(\B x_1, \B x_2, \B x_3)$,
with any of the regularizations \re{6.1}--\re{6.4},  follows from the symmetries 1--3, namely  that
it actually depends only on the three inter-point distances $x_{12}, x_{13}, x_{23}$, where
$x_{ij} = |\B x_{ij}|$. Explicitly, this is readily seen if the factors
$\ds \frac{e^{-\mu |\B v + \B x_{ij}|}}{|\B v + \B x_{ij}|}$ in equations \re{6.1}--\re{6.4} are expanded in
spherical harmonics ($\mu \equiv 0$ in \re{6.3}, \re{6.4}), the angular integrations
$\int \D{}{\hat{\B v}} ... ~$ are
carried out, and the orthogonality properties of the spherical harmonics are used, then (after the
remaining integration over $\D{}v$), the result is seen to depend only on the lengths of the two vectors $\B x_{12}$,
$\B x_{13}$ and on the angle between them (or, equivalently, on $x_{12}, x_{13}, x_{23}$).

    The direct calculation of the regularized potential,
with any of the regularizations \re{6.1}--\re{6.4}, is complicated.
Instead, we propose a representation for the function \re{5.9} in which
its dependence on scalar arguments is manifest.
This greatly simplifies the regularization and evaluation of $U$.
Let us apply the well known formula:
$$
\frac1r=\frac1{\sqrt{\pi}}\int^{\infty}_{-\infty}\!\D{} k {\rm e}^{-k^2r^2}
$$
to each factor of the integrand of the expression \re{5.9}
(which shall be treated formally). Then changing the order of integration we have:
%
\begin{eqnarray}\lab{6.5}
U(\B x_1, \B x_2, \B x_3)&=-\frac1{\pi^{3/2}}
\inta\D3k\inta\D3z\,{\rm e}^{-k_1^2(\B z-\B x_1)^2-k_2^2(\B z-\B x_2)^2-k_3^2(\B z-\B x_3)^2}\nn\\
&=-\inta\frac{\D3k}{k^3}\,{\rm e}^{-(k_1^2k_2^2x_{12}^2+k_2^2k_3^2x_{23}^2+k_1^2k_3^2x_{13}^2)/k^2}\nn\\
&=-\inta\D{}{\hat{\B k}}\int\limits^{\infty}_0\frac{\D{}k}{k}\,
{\rm e}^{-(\hat k_1^2\hat k_2^2x_{12}^2+\hat k_2^2\hat k_3^2x_{23}^2+\hat k_1^2\hat k_3^2x_{13}^2)k^2}
\end{eqnarray}
where $\hat{\B k}=\B k/k$. It is obvious in this form that $U(\B x_1, \B x_2, \B x_3)=U(x_{12}, x_{23}, x_{13})$
and, in addition, that the internal integral in the last line of \re{6.5} is divergent at
its lower boundary $k=0$.

    The potential difference:
%
\begin{equation}\lab{6.6}
U(x_{12}, x_{23}, x_{13})-U(y_{12}, y_{23}, y_{13})=-\inta\D{}{\hat{\B k}}\!\int\limits^{\infty}_0\frac{\D{}k}{k}\,
\left[{\rm e}^{-X^2k^2}\!-{\rm e}^{-Y^2k^2}\right],
\end{equation}
where $X^2=\hat k_1^2\hat k_2^2x_{12}^2+\hat k_2^2\hat k_3^2x_{23}^2+\hat k_1^2\hat k_3^2x_{13}^2$,
$Y^2=\hat k_1^2\hat k_2^2y_{12}^2+\hat k_2^2\hat k_3^2y_{23}^2+\hat k_1^2\hat k_3^2y_{13}^2$,
will be finite since infinite constants $U_0$ (see \re{5.10}) from the first and second terms of \re{6.6}
mutually cancel. Indeed, using the cut-off parameter $\varepsilon$ in
the internal integral in r.h.s. of \re{6.6} yields:
$$
\int\limits^{\infty}_\varepsilon\frac{\D{}k}{k}\,
\left[{\rm e}^{-Y^2k^2}-{\rm e}^{-X^2k^2}\right]=
\left[\int\limits^{\infty}_{Y\varepsilon}-\int\limits^{\infty}_{X\varepsilon}\right]
\frac{\D{}t}{t}\,{\rm e}^{-t^2}=
\int\limits^{X\varepsilon}_{Y\varepsilon}\frac{\D{}t}{t}\,{\rm e}^{-t^2}
\mathop{\longrightarrow}\limits_{\varepsilon\to0}\ln\frac{X}{Y},
$$
i.e., the integral is convergent.

    Next, we introduce angular variables $\{\vartheta,\varphi\}$ on the unit sphere in $k$-space, so that
$\hat k_1=\sin\vartheta\cos\varphi$, $\hat k_2=\sin\vartheta\sin\varphi$, $\hat k_3=\cos\vartheta$. Then
%
\begin{equation}\lab{6.7}
\fl\qquad\qquad
U(x_{12}, x_{23}, x_{13})-U(y_{12}, y_{23}, y_{13})=W(\bar x_{12},\bar x_{23}, \bar x_{13})-
W(\bar y_{12}, \bar y_{23}, \bar y_{13}),
\end{equation}
where
%
\begin{eqnarray}\lab{6.8}
\fl
W(\bar x_{12},\bar x_{23}, \bar x_{13})=\ha\int\limits_0^{2\pi}\!\D{}\varphi\!
\int\limits_0^{\pi}\!\sin\vartheta\,\D{}\vartheta
\ln\left[(\bar x_{12}\sin\vartheta\cos\varphi\sin\varphi)^2 + (\bar x_{23}\cos\vartheta\sin\varphi)^2
\right.\nn\\
\hspace{46ex}
\left.{}+ (\bar x_{13}\cos\vartheta\cos\varphi)^2\right]
\end{eqnarray}
and $\bar x_{ij}=x_{ij}/a$~. The arbitrary constant $a$ (with dimension of length) is introduced
so that the argument of the logarithm will be dimensionless. Actually, the potential difference \re{6.7}
does not depend on $a$ while the function \re{6.8} itself does. Since this function is well defined and finite,
it can be considered, up to some additive constant, as the regularized potential:
%
\begin{equation}\lab{6.9}
\tilde U(\B x_1, \B x_2, \B x_3)=W(\bar x_{12},\bar x_{23}, \bar x_{13})-W_0.
\end{equation}
The choice of the constant $W_0$ is a matter of taste; it can be canceled by an
appropriate rescaling of the constant $a$: $W(x_{12}/a,...)=W(x_{12}/b,...)+4\pi\ln(b/a)$.
Thus an arbitrariness of the regularized potential arises due to the scale constant $a$.

    We note that the regularized function \re{6.9} obeys the following scaling property:
\medskip

\noindent
($\widetilde{\mbox{iv}}$)
scale invariance: $\tilde U(\lambda\B x_1, \lambda\B x_2, \lambda\B x_3) =
\tilde U(\B x_1, \B x_2, \B x_3)+4\pi\ln\lambda$, where $\lambda\in\Bbb R_+$.
\medskip

    The inner integration (over $\vartheta$) in \re{6.8} can be performed explicitly.
Then the change of variable $\varphi\to s=\cos\varphi$ yields:
%
\begin{equation}\lab{6.10}
\tilde U(\B x_1, \B x_2, \B x_3)=4\pi\ln\frac{x_{13}+x_{23}}{4a} + I(\xi,\eta),
\end{equation}
where
%
\begin{eqnarray}
I(\xi,\eta)=4\int\limits_{-1}^{1}\frac{\D{}s}{\sqrt{(s+\xi)^2+\eta^2}}
\,\mathrm{arctan}\sqrt{\frac{(s+\xi)^2+\eta^2}{1-s^2}},
\lab{6.11}\\
\xi=\frac{x_{13}^2-x_{23}^2}{x_{12}^2},\qquad
\eta^2=\frac{[(x_{13}+x_{23})^2-x_{12}^2][x_{12}^2-(x_{13}-x_{23})^2]}{x_{12}^4},
\lab{6.12}
\end{eqnarray}
and we have chosen for convenience: $W_0=4\pi(\ln2-1)$. We note that the interparticle distances
must satisfy the triangle inequalities: $x_{13}+x_{23}\ge x_{12}$, $x_{23}+x_{12}\ge x_{13}$ and
$x_{12}+x_{13}\ge x_{23}$.

    The regularized potential \re{6.10}--\re{6.12} possesses the permutational invariance (iii)
implicitly.
This is evident from the fact that any particle permutation is equivalent to some renumbering of
$k$-variables in the integrals \re{6.5}, \re{6.6} and, finally, to another choice of angular
variables in the integral \re{6.8}.

    In the particular cases where the points $\B x_1$, $\B x_2$ and $\B x_3$ lie on a straight line
the integral \re{6.11} can be calculated analytically:
%
\begin{equation}\lab{6.13}
\tilde U(\B x_1, \B x_2, \B x_3)=4\pi\ln\frac{x_>}{2a},
\quad\mbox{where}\quad x_>={\rm max}(x_{12},x_{13},x_{23}).
\end{equation}

Another analytically solvable case is that of equidistant points, $x_{12} = x_{13} = x_{23} = r$,
whereupon in \re{6.12}, $\xi = 0$ and $\eta^2 =3$, so that $I$ of \re{6.11} is a finite constant
independent of $r$. Thus $\tilde U (r,r,r) = 4 \pi \ln (r/a) + c_1$, where $c_1$ is a finite constant,
which we can ignore (it does not affect energy differences).
For convenience we shall use ``atomic units", that is energies will be in units of $m \alpha^2$,
and lengths in units of $a = m \alpha$, where $\ds \alpha = \frac{g^2}{16 \pi m^2}$ is the dimensionless
 ``fine structure constant". The total potential $\ds V = V_{33}^{(2)} + V_{33}^{(3)}$ (cf. eqs. \re{5.7}
 and \re{5.8}) is (with $\lambda = 0$), in atomic units,
%
\begin{equation}  \lab{6.14}
V(r) = - \frac{3}{r} + \gamma \ln r , ~~~~ (r ~{\rm is}~ r m \alpha,~ {\rm and}~ V ~{\rm is}~ V / m \alpha^2)
\end{equation}
where $\ds \gamma = 4 \kappa/g$.
We see that $V(r)$, in this equidistant-points subspace, is a
uniformly increasing, logarithmically confining potential (for $\gamma >0$). Note that
$\ds V(r) \simeq - \frac{3}{r}$ for small $r$ ($r \to 0_+$) but $ V(r) \simeq \gamma  \ln r$
for large $r$. Recall that if $\kappa = \gamma = 0$, the bound-state eigenvalue spectrum (in
atomic units) is the
Rydberg spectrum $\ds \epsilon_n = - \frac{3}{2} \frac{1}{n^2}$, where $n = 1, 2, 3, ...$,
and there are no bound states
for $\epsilon >0$. However, for $\gamma >0$, the logarithmic confining potential stretches out this
Rydberg spectrum, so that there is a purely bound-state spectrum for $\epsilon > 0$.
Using various approximations \cite{Q-R79,S-B08} one can estimate $\epsilon_n \simeq \gamma\ln n$ for
$n\gg1$. (The repulsive contact (delta-function) potentials, which we have ignored by taking $\lambda =0$,
are of little consequence, since such repulsive contact potentials have an insignificant effect on the
energy spectrum.)
\medskip
\par
Other regularization methods lead, basically,  to  the same results. For example, if we use the
cut-off regularization of \re{6.3}, then for the case $x_{13} = 0$, we obtain
$\Ub_R = - 4 \pi [ 1 + \ln(R/x_{12})] = 4 \pi \ln(x_{12}/a) - c_2$ , where $a$ is an arbitrary
length parameter (length unit), and $c_2 = 4 \pi [1 + \ln (R/a)]$ is a very large constant, which
has to be absorbed into a redefined (shifted) energy, as in \re{5.12}. This result is the same as
eq. \re{6.13}.

    In the general case, a numerical integration of \re{6.11} is required.
We illustrate the behavior of the potential in Figure 1
for the particular case $\B x_1 = \B a$,  $\B x_2 = -\B a$ as a
function of $\B x_3 = \B r$. The value of potential for arbitrary
configuration can be obtained from it using the symmetry properties
(i)--(iii) and ($\widetilde{\mbox{iv}}$).
%
%
\begin{figure}[htb]
\begin{center}
\includegraphics[scale=0.8]{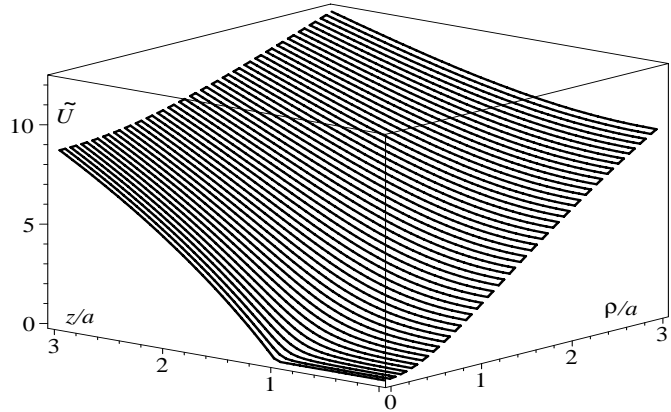}
\caption{The potential $\tilde U(\B a, -\B a, \B r)$ as a
function of $\B r=\{x,y,z\}$; $\rho=\sqrt{x^2+y^2}$; $a = |\B a|$.
The function is symmetric under the inversion $z\to-z$ and rotation around $0z$.
In particular, $\tilde U=4\pi\theta(|z|-a)\ln\ha(|z|/a+1)$ if $\rho=0$.}
\end{center}
\end{figure}
\par
    In the case where one of the points is far from the others, the equality
\re{6.13} is valid asymptotically. Thus the regularized potential
reveals logarithmic confinement properties.

    A detailed analysis of the (non-relativistic) bound-state spectrum for the general case
requires the solution of the three body equation \re{5.6}. This is a quite challenging task
in itself. However, from the confining nature of the three-point potential, we can see
that the spectrum will reflect confinement, much like for the equidistant-points subspace
of \re{6.14}.


\section{Concluding remarks}

    We have considered generalizations of the Wick-Cutkosky (massless
scalar Yukawa) model that include nonlinear mediating fields.
Covariant Green functions were used to eliminate the mediating
field, thus arriving at a Lagrangian that contains nonlocal
interaction terms.

In the case of a massless mediating field $\chi$,
with a $\th\kappa\chi^3{+}\qu\varkappa\chi^4$ nonlinearity,
we evaluate the corresponding interaction term explicitly
and show that the kernel has the form of a three- and four-point
``cluster potential", cf. \re{4.4}, \re{4.5}.

    We consider the quantized version of this model in the Hamiltonian formalism, and
use the variational method, with trial states built from Fock-space
components, to derive a relativistic integral wave equation for the
three-particle system. The kernels (relativistic potentials) are
shown to contain one-quantum exchange terms and a three-point
cluster term. In the non-relativistic limit we evaluate the explicit
coordinate-space form of the interaction potentials and show that
they consist of attractive pairwise Coulombic potentials and a
cluster three-point confining potential. The three-point potential,
which arises from the  $\frac{1}{3} \kappa \chi^3$ term in the
Hamiltonian, is divergent (and so needs regularization), but the
potential differences are finite. The regularized three-point
potential is shown to be logarithmically confining, and dependent
only on the three inter-point distances. Its evaluation, for
arbitrary values of its arguments, is shown to be reducible to a
single quadrature.

The three-body wave-equation derived in this paper is quite complicated and must be
solved using approximation methods. This will be the subject of forthcoming work.

The three-particle trial state \re{5.1} is found to be the simplest
variational ansatz which manifests the
confinement properties of the model.
However, other sectors of the Fock
space in the variational problem are also of interest.
For example,
an open problem is the role of the three-point interaction
in the particle-antiparticle problem.
It was pointed out in the section 5 that
the simple variational particle-antiparticle trial state
$|1{+}\bar1\rangle$ does not sample the $H^{(3)}_{\rm int}$ term
\re{4.7} of the Hamiltonian. Thus, this term does not
influence the variational wave equation
derived by using only $|1+\bar1\rangle$  (see \cite{Dar00,D-D04,ERD05}),
in which case the
only Coulomb-like interaction arise.
But the inclusion of both the
$|1{+}\bar1\rangle$ and $|2{+}\bar2\rangle$ sectors
leads to a coupled set of two many-body wave-equations  \cite{ERD06}
in which the effects of $H_{\rm int}^{(3)}$ and $H_{\rm int}^{(4)}$ are present.
Whether these effects are confining is a question that needs to be investigated.

Lastly, we comment on the role of ``chion'' Fock-space sector
in the variational bound state problem
within the reduced Hamiltonian formalism of QFT  used in this work.
This role can be examined by
taking into account the $\chi_0$-dependent
extra terms $\Delta{\cal L}$
of the total non-local Lagrangian \re{3.4}.
They are at least quadratic in $\chi_0$
including the free-field term
$\ha\partial_\mu\chi_0\partial^\mu\chi_0$ and interaction terms;
see eq.\re{3.7}.
Thus the additional Hamiltonian corresponding to the extra terms,
$\Delta H$,  has no
effect on variational states $|\Psi\rangle$ without free ``chions''
(i.e., quanta of the field $\chi_0$), since
$\langle\Psi|\Delta H|\Psi\rangle=0$ for such states.
A non-trivial contribution to
a variational bound-state problem may arise from states with two or more
virtual ``chions'' but this is a higher-order effect in the coupling constants
($\kappa$, $\varkappa$ or others) of the potential ${\cal V}$.


\ack
The authors are grateful to
V. Tretyak,
T. Krokhmalskii and Yu.~Yaremko for helpful discussion of this work.


\section*{Appendix. Nonlocal Lagrangian from a nonlinear dipole model}
\renewcommand{\theequation}{A.\arabic{equation}}
\setcounter{equation}{0}

    In this section we consider a model which is built in
analogy to the linear ``dipole model" \cite{Bla74,D-D04} that
simulates the confinement interaction of quarks in mesons. This
model is nonlinear and gives Yukawa~+~cluster interactions.
It is specified by the Lagrangian
%
\begin{equation}\lab{A.1}
{\cal L} =\partial_\mu\phi^*\partial^\mu\phi - m^2\phi^*\phi - \qu\lambda(\phi^*\phi)^2 +
\rho \, (\chi + \ha\varphi) +
\partial_\mu\chi \, \partial^\mu\varphi - {\cal V}(\varphi),
\end{equation}
where both the $\chi(x)$ and $\varphi(x)$ are real massless scalar
fields and $\rho = - g \phi^* \phi$ as in \re{3.1}.

    The variation of the action \re{2.1}, \re{A.1} leads
to the coupled set of the Euler-Lagrange equations,
%
\begin{eqnarray}
& (\square + m^2)\phi = - g \, \phi \, (\chi+\ha\varphi) - \lambda\phi(\phi^*\phi), \lab{A.2}\\
& (\square + m^2)\phi^* = - g \, \phi^* \, (\chi+\ha\varphi) - \lambda\phi^*(\phi^*\phi), \lab{A.3}\\
& \square\varphi = \rho, \lab{A.4}\\
& \square\chi = \ha\rho - {\cal V}\,'(\varphi), \lab{A.5}
\end{eqnarray}
which determine the field dynamics.

    Equations \re{A.4} and \re{A.5} possess the exact formal solution:
%
\begin{eqnarray}
& \varphi = D*\rho, \lab{A.6}\\
& \chi = D*\left\{\ha\rho-{\cal V}\,'(\varphi)\right\} =
D*\left\{\ha\rho-{\cal V}\,'(D*\rho)\right\}, \lab{A.7}
\end{eqnarray}
which can immediately be used in the r.h.s. of eqs. \re{A.2},
\re{A.3}:
%
\begin{equation}\lab{A.8}
(\square + m^2)\phi = - g\phi D*\left\{\rho-{\cal V}\,'(D*\rho)\right\} - \lambda\phi(\phi^*\phi),
\end{equation}
and similarly for $\phi^*$. These equations can be derived from
$\delta \, I = 0$, with a Lagrangian identical to \re{3.6} (but note
that no iterative expansion, like that in eq. \re{3.6}, needs to be
made in this case).


\end{document}